\documentclass[12pt]{article}

\usepackage{amssymb}
\usepackage{amsmath}
\usepackage{latexsym}
\usepackage{yfonts}

\catcode`@=11
\renewcommand{\section}{\@startsection{section}{1}{0pt}{\medskipamount}
{\medskipamount}{\large\bf}} \numberwithin{equation}{section}
\catcode`@=12

\def\beq{\begin{eqnarray}}    
\def\eeq{\end{eqnarray}}      

\def\ln{\,\mbox{ln}\,}                  
\def\str{\,\mbox{str}\,}                

\def\pa{\partial}                       

\def\={\ =\ }

\begin{document}

\begin{center}

{\Large\bf General quantum-mechanical setting for field-antifield
formalism as a hyper-gauge theory}

\vspace{18mm}

{\large Igor A. Batalin$^{(a,b)}\footnote{E-mail:
batalin@lpi.ru}$\;, Peter M. Lavrov$^{(b, c)}\footnote{E-mail:
lavrov@tspu.edu.ru}$\; }

\vspace{8mm}

\noindent ${{}^{(a)}}$
{\em P.N. Lebedev Physical Institute,\\
Leninsky Prospect \ 53, 119 991 Moscow, Russia}

\noindent  ${{}^{(b)}} ${\em
Tomsk State Pedagogical University,\\
Kievskaya St.\ 60, 634061 Tomsk, Russia}

\noindent  ${{}^{(c)}} ${\em
National Research Tomsk State  University,\\
Lenin Av.\ 36, 634050 Tomsk, Russia}

\vspace{20mm}

\begin{abstract}
\noindent
A general quantum-mechanical setting is proposed  for the field-antifield
formalism as a unique hyper-gauge theory in the field-antifield  space.
We formulate a  Schr\"{o}dinger-type equation to describe the quantum evolution
in a "current time" purely formal in its nature.  The corresponding
Hamiltonian is defined in the form of a supercommutator of the
delta-operator with a hyper-gauge Fermion.  The initial wave function is
restricted to be annihilated with the delta-operator. The Schr\"{o}dinger's
equation is
resolved in a closed form of the path integral, whose action contains the
symmetric  Weyl's  symbol of the Hamiltonian.  We take the path integral
explicitly in the case of being a hyper-gauge Fermion an arbitrary function rather
than an operator.
\end{abstract}

\end{center}

\vfill

\noindent {\sl Keywords:} field-antifield formalism,
quantum antibracket, Weyl symbol
\\

\noindent PACS numbers: 11.10.Ef, 11.15.Bt
\newpage

\section{Introduction}

The field-antifield (BV) formalism \cite{BV,BV1} is known as the most powerful method
for covariant (Lagrangian) quantization of gauge-field theories of the
general kind, with  general  open gauge algebra,  both irreducible  or any-stage
reducible. On the other hand, the field-antifield formalism was derived
directly from the Hamiltonian generalized  canonical quantization
\cite{FVh1,FVh2,BVh,BFr1,BF2},  and
thus, the physical unitarity was guaranteed by construction.

The main ingredient of the field-antifield formalism is  the quantum master
equation  formulated in terms of the nilpotent  odd Laplacian also known as
the delta-operator.  The nilpotency  of the delta-operator causes the
natural arbitrariness \cite{BBD,BLT-BV,BLT-EPJC}
for the quantum master action. That quantum
arbitrariness is realized in the form of the so-called anticanonical master
transformations \cite{BLT-EPJC}, infinitesimal or finite \cite{BV,BLT-EPJC}.
These master transformations do generalize comprehensively the famous
BRST transformations
\cite{brs1,t}.
Gradually, it was realized  that  the whole field-antifield formalism has
the characteristic  features  of some unique hyper-gauge theory which lives
in  the antisymplectic phase space. These ideas take their most symmetric form
within the framework of the so-called $W$-$X$ construction proposed recently
\cite{BBL}.

In the present paper, we would like to make a new step to unique
hyper-gauge theory.  Namely,  we would like to propose a new
quantum-mechanical setting  for the field-antifield formalism as a unique
hyper-gauge theory.
First,  we introduce a new "current time",  purely formal in its nature.
Then, we define the quantum evolution in the new  "current time" by means of the
Schr\"{o}dinger equation with a  Hamiltonian chosen as a general delta-exact
form, which is a supercommutator of the delta-operator, with some hyper-gauge
Fermion.  That delta-exact form is directly related to the pair of dual
quantum antibrackets \cite{BM1,BM2}. Quantum state is described in terms of a wave
function  living  in the field-antifield  "configuration space".   We
restrict the initial wave function to satisfy the quantum master equation,
i. e. to be annihilated with the delta-operator.  Because of the
delta-exactness of the Hamiltonian, it follows immediately that the current
state satisfies  the quantum master equation, as well.  In this way, a
hyper-gauge Fermion describes the natural
arbitrariness in resolving the quantum master equation \cite{BBD,BLT-BV,BLT-EPJC}.

We resolve the Schr\"{o}dinger equation in a closed form,  in  terms of the
path integral whose action contains the corresponding symbol of the
Hamiltonian
operator. In the previous paper \cite{BL}  we have used the normal  $ZP$
symbol, while in the present paper we make use of the symmetric Weyl's
symbol. Of course, it is the simplest possible case when a hyper-gauge Fermion is an
arbitrary function, rather than an operator.   In that case, in the path
integral, the momenta integration yields the delta-functional concentrated  on the
orbit of the anticanonical transformation generated by the hyper-gauge
Fermion. When resolving the delta-functional, the latter yields the corresponding
Jacobian. If one uses the normal $ZP$ symbol \cite{Ber1}  then the latter
Jacobian equals to one \cite{BL}, while in the case of the Weyl's symbol, the corresponding
Jacobian is rather nontrivial. The difference between the two cases is
caused by the fact
that the boundary conditions, the integration trajectory should satisfy to,
appears dependent actually of the type of the symbol chosen.  In the present
article, we have calculated explicitly the Jacobian yielded by the
delta-functional  in the case of the symmetric Weyl's symbol chosen.
\\

\section{Operators and symbols}

Let  $Z^{A}$ be the complete set of field-antifield antisymplectic
variables, and let $P_{A}$ be their canonically conjugate momenta,
\beq
\label{Q1}
 P_{A} =:  - i  \hbar  \rho^{ - 1/2}  \partial_{A}  (-1)^{
\varepsilon_{A} } \rho^{ 1/2},  \quad  [ Z^{A},  P_{B} ]  =  i  \hbar
\delta^{A}_{B},   
\eeq
where the Grassmann parities of the operators in (\ref{Q1}) are denoted by
\beq
\label{Q2}
\varepsilon_{A} =:  \varepsilon( Z^{A} ) = \varepsilon( P_{A} ) . 
\eeq
All these operators are Hermitian with respect to the standard
scalar product
\beq
\label{Q3}
<\psi| \phi >   =:   \int  d\mu( Z ) \psi^{*}( Z ) \phi( Z ),  \quad   d\mu(
Z )  =:   dZ  \rho( Z ).     
\eeq
Let us proceed with the Cauchy problem for the Schr\"{o}dinger equation
\beq
\label{Q4}
i \hbar \;\!\partial_{ t } \Psi( t )   =  H( Z, P ) \Psi( t ),  \quad
\Psi( t =  0 )  =  \Psi_{0},   
\eeq
where the Hamiltonian is defined as
\beq
\label{Q5}
H( Z, P )  =:  ( i \hbar )^{-1} [ \Delta( Z, P ),  F( Z, P ) ],
\eeq
\beq
\label{Q5a}
( i \hbar )^{-1}  {\rm ad}( H )   =
( i \hbar )^{-2}  \frac{2}{3}  (  {\rm ad}_{ \Delta }( F ) +
{\rm ad}_{ F }( \Delta )  ),   
\eeq
\beq
\label{Q5b}
[ {\rm ad}_{ \Delta }( F ),  {\rm ad}_{ F }( \Delta ) ]   =
 -  \frac{1}{4}  {\rm ad}( \Delta ) \; {\rm ad}_{ \Delta }(\frac{1}{2}[ F,  F ]  ),  
\eeq
with ${\rm ad}_{ \Delta }( F )$ being  the adjoint action of the
$\Delta$-generated quantum  $F$-antibracket \cite{BM1,BM2},  and the nilpotent
Fermion Hermitian  operator $\Delta$  being  defined as
\beq
\label{Q6}
\Delta( Z, P )  =:  \frac{1}{2} \rho^{ - 1/2 } P_{A} \rho E^{AB}
P_{B} \rho^{ - 1/2 } (-1)^{ \varepsilon_{B} }  +  ( i \hbar )^{2}
\nu(Z),    
\eeq
with $E^{AB}( Z )$ being the antisymplectic metric and $\nu( Z )$ being
the Fermion function \cite{GL,BB,BB1} introduced in order to provide for the measure
density $\rho$ to be independent of the antisymplectic metric $E^{AB}(Z)$.
In  terms of the operator (\ref{Q6}), the initial state $\Psi_{0}( Z )$
in the second in (\ref{Q4}) is restricted to satisfy the quantum master
equation
\beq
\label{Q7}
\Delta( Z, P ) \Psi_{0}  =  0,     
\eeq
which implies the same equation as to the current state $\Psi( t, Z)$,
\beq
\label{Q8}
\Delta( Z, P ) \Psi( t )  =  0,   
\eeq
due to the property
\beq
\label{Q9}
[ \Delta( Z, P ),  H( Z, P ) ]  =  0.    
\eeq
In turn, we identify
\beq
\label{Q9a}
\Psi_{0}( Z )  =  \exp\left\{  \frac{ i }{ \hbar }  W( Z ) \right\},  \quad
\Psi(1, Z  ) =
\exp\left\{ \frac{ i }{ \hbar } W'( Z ) \right\}   
\eeq
with $W( Z )$ and $W'( Z )$ being the original
and the new (transformed) master action, respectively.

The operator $F( Z, P )$ in (\ref{Q5}) is an arbitrary  Fermion  Hermitian
operator.  Its  arbitrariness  describes the one of the
field-antifield formalism as a hyper-gauge theory.  In that sense,
one can consider the condition (\ref{Q7})/(\ref{Q8} ) as the one as to define
physical states.

Of course, the current time $t$ in (\ref{Q4}) is purely formal in its nature.
However,  one is allowed to use it  formally in all aspects of
quantum description in the usual way. The situation here resambles
a bit the one with the proper time of Schwinger/ Fock \cite{Sch,Fock}.   For
instance, by proceeding from the  Schr\"{o}dinger's  picture (\ref{Q4}) one
can easily change for  the Heisenberg's  or  the  Dirac's  picture,
if  desired  for the  sake  of  technical  convenience.

With respect to the operator-valued functions of the basic elements
$Z^{A}$ and $P_{A}$, one can change for the corresponding symbol
calculus of  Berezin \cite{Ber1},  such as  $ZP$-normal symbols,  or
symmetric  symbol  of Weyl, and so on. In our previous consideration
\cite{BL}, the corresponding formalism has been developed for $ZP$
normal symbol, the simplest one technically.  Here, we will consider
below the formalism based on the use of  symmetric  symbols of  Weyl.

Given an operator $H( Z, P )$ in the symmetric  Weyl's  form,
\beq
\label{Q11}
H( Z, P )  =:   \left(  \exp\left\{  Z^{A} \frac{ \partial }{ \partial {\bar Z}^{A}}
+
P_{A} \frac{ \partial }{ \partial {\bar P}_{A}} \right\}    H( \bar{Z}, \bar{P} )
\right) \Big|_{ \bar{Z}  =  0, \bar{P}  =  0 },     
\eeq
a function  $H ( \bar{Z}, \bar{P} )$ of classical phase variables $\bar{Z}^{A},
\bar{P}_{A}$ is called a  Weyl's  symbol.  In what follows below
we will use the short-hand notation for Weyl's symbols $H( Z, P )$ as
functions of classical phase variables $Z^{A}, P_{A}$.

It follows from (\ref{Q11}) that the star multiplication  for Weyl's symbols has
the form
\beq
\label{Q12}
\star  =:  \exp\left\{  \frac{ i \hbar }{ 2 }  \left(  \frac{\overleftarrow{\partial }}{
\partial Z^{A} }  \frac{ \overrightarrow{\partial }}{ \partial P_{A} }  -
\frac{\overleftarrow{\partial }}{ \partial P_{A} } ( -1)^{ \varepsilon_{A} }
\frac{ \overrightarrow{\partial } }{ \partial  Z^{A}  }   \right)  \right\}.    
\eeq
In terms of (\ref{Q12}),  the operator valued definition  (\ref{Q5})
and the property (\ref{Q9}) rewrite for symbols as
\beq
\label{Q12a}
H( Z, P )  =  ( i \hbar )^{-1} [ \Delta( Z, P ),  F( Z, P ) ]_{ \star }\;\!, \quad
[ \Delta( Z, P ),  H( Z, P ) ]_{ \star }  =  0\;\!,    
\eeq
where  $[ \;, \;]_{ \star }$ means the symbol  supercommutator
\beq
\label{Q12b}
[ A( Z, P ),  B( Z, P ) ]_{ \star }   =
:   A( Z, P )  \star  B( Z, P )   -   B( Z, P )  \star  A( Z, P )
(-1)^{ \varepsilon( A ) \varepsilon( B ) } .   
\eeq

\section{Path integral resolution for Schr\"{o}dinger
equation in terms of Weyl's symbols}

Given the Weyl's symbol $H( Z, P )$ of the Hamiltonian, a formal solution  to
the Cauchy problem (\ref{Q4}) is
\beq
\label{Q13}
\Psi( 1, Z )  = ( 2  \pi  i  )^{ - D }
\int dY dP\; U\left( 1, \frac{1}{2} ( Z + Y ) , P \right)
\exp\left\{
\frac{ i }{ \hbar }P\;\!( Z  -  Y ) \right\} \Psi_{ 0 }( Y ),    
\eeq
where $D$   is  the  number of  Bosons  among   $Z^{A}$,
and $U( t, Z, P )$ is the symbol of the evolution operator,
\beq
\label{Q13a}
i  \hbar \partial_{ t } U( t, Z, P )  =  H( Z, P ) \star  U( t, Z, P)\;\!, \quad
U( 0, Z,P )  =  1\;\!,  
\eeq
with  $H( Z, P )$ being  the symbol  (\ref{Q12a}) of the Hamiltonian.
By making use of the standard
functional methods \cite{Ber1,BSh,Fradkin},  one derives the
following path integral representation for $ U( 1, Z, P )$ (see Appendix A for details)
\beq
\label{Q14}
 U( 1, Z, P ) =  \left\langle
 \exp\left\{-\frac{ i }{ \hbar } P_{A} (  Z^{A}( 1 )  -  Z^{A}( 0 )  )  - \frac{ i }{ \hbar }
\int_{0}^{1}  dt \;\! H( Z( t
), P( t ) )  \right\}  \right\rangle,  
\eeq
where the average is defined by
\beq
\label{Q15}
\langle ( ... )  \rangle  =:  \frac{  \int \mathcal{D} V  \mathcal{D} P  ( ... )
\exp\left\{
\frac{ i }{ \hbar }   \int_{0}^{1}  dt  P_{A} \dot{Z}^{A}  \right\}  }
{  \int  \mathcal{D} V  \mathcal{D} P
\exp\left\{  \frac{ i }{ \hbar }  \int _{0}^{1}
dt  P_{A} \dot{Z}^{A}  \right\}  }\;\!, 
\eeq
the integration  trajectory $Z^{A}( t )$  is restricted to satisfy the Weyl's
boundary condition
\beq
\label{Q16}
Z^{A}( 1 )  +  Z^{A}( 0 )  =  2  Z^{A},    
\eeq
which resolves in terms of unrestricted integration  velocities $V^{A}( t )$,
\beq
\label{Q17}
Z^{A}( t )  =:   Z^{A}  +  \int_{0}^{1}  dt' \;\! \frac{1}{2} {\rm sign}(  t - t'  )
V^{A}( t' ),   \quad   \dot{Z}^{A}( t )  =  V^{A}( t ).      
\eeq

It is also worth to mention that the famous  Berezin's  formula
(\ref{Q13}) has a nice interpretation within the following basic
proposal related directly to the symbol multiplication law,
\beq
\label{Q17n}
\Psi( 1, Z )  =  \int  d Y  {\cal K}( 1, Z, Y )  \Psi_{ 0 }( Y ),    
\eeq
\beq
\label{Q17a}
{\cal K}( 1, Z, Y )= ( 2  \pi  i )^{ - D }  \int  dP \; U( 1, Z, P )  \star
\exp\left\{
\frac{ i }{ \hbar }  2 P  ( Z  -  Y )  \right\}.
\eeq
By inserting the standard  Weyl's  multiplication  for the
$\star$ (\ref{Q12}), we get
\beq
\label{Q17b}
{\cal K}( 1, Z, Y )   =  \int   dX  ( 2 \pi  i  )^{ - D }  dP\;  U(  1,  Z  -  X,  P
) \; \delta(  2  X  -  Z  +  Y  )  \exp\left\{  \frac{ i }{ \hbar } 2 P X \right\} .
\eeq
It  follows immediately from (\ref{Q17b}) that the standard  Berezin's  formula
(\ref{Q13}) holds,
\beq
\label{Q17c}
{\cal K}( 1, Z, Y )  =  ( 2  \pi  i  )^{ - D }
\int  dP\; U\left( 1, \frac{ 1 }{ 2 }( Z  +  Y ),  P \right)
\exp\left\{\frac{ i }{ \hbar } P( Z  -  Y ) \right\},  
\eeq
together with its inverse \cite{Ber1},
\beq
\label{Q17d}
U( 1,  Z,  P )   =   \int   dX   {\cal K}\left(1,  Z  +  \frac{1}{2}  X,   Z  -
\frac{1}{2}  X \right)
\exp\left\{  -  \frac{ i }{ \hbar }  P X  \right\}.   
\eeq

For details as to how the formula (\ref{Q17c}) does  follow  from the general
definition of the Weyl's operators, see Apppendix B.

\vspace{0.1cm}

\section{ Taking the Weyl's path integral in the
simplest case}

For the sake of further simplicity, here we choose the Darboux co-ordinates,
\beq
\label{Q18}
E^{AB}  = {\rm const}( Z ),    \quad \rho( Z ) = 1,  \quad  \nu( Z )  =  0.   
\eeq
Let us consider the simplest case, when the hyper-gauge Fermion
$F$ is an arbitrary function of $Z$,
\beq
\label{Q19}
F  =  F( Z ) \;\Rightarrow\; H( Z, P )  = P_A ( F,  Z^{A} ). 
\eeq
Then, the $P$-integration yields the delta functional
\beq
\label{Q20}
\delta[ \dot{Z}^{A}  -  ( F, Z^{A} ) ] =
J^{-1}[ Z ] \; \delta[ Z^{A} - Z_{R}^{A} ] \;  ,    
\eeq
concentrated on the orbit $Z_{R}^{A}( t )$ of an anticanonical transformation,
 with the $F( Z )$
being a generator,
\beq
\label{Q21}
\dot{Z}^{A}  =  ( F, Z^{A} ).   
\eeq
By resolving the latter together with (\ref{Q16}), we get the solution
\beq
\label{W7}
Z_{R}^{A}( t )   =:
\frac{  \exp\{  t \;\! {\rm ad}( F )  \}  }{   \exp\{  {\rm ad}( F )  \}  +  1  }\;
2  Z^{A}=
\frac{  \exp\left\{  ( t - \frac{1}{2} ) {\rm ad}( F )  \right\} }{{\rm cosh}\left(  \frac{1}{2}
{\rm ad}( F )  \right)  } \; Z^{A}, 
\eeq
where
${\rm cosh}( x )$
is the ordinary hyperbolic cosine.
In a short-hand matrix notation,  the logarithm of the delta-functional's Jacobian
is expressed as:
\beq
\label{W8}
\ln J[Z]  =  -  \int_{0}^{1}  d\lambda   \int_{0}^{1}  dt
\str[  \Gamma( t,  t )  X( t )  ],    
\eeq
where  $\Gamma( t, t' )$ satisfies the equation
\beq
\label{W9}
[ \partial_{ t }  1  - \lambda  X( t ) ]  \Gamma( t,  t' )   =
\delta( t  -  t' )  1,    
\eeq
and the boundary condition
\beq
\label{W10}
\Gamma(  t  =  1,  t'  )  +  \Gamma(  t  =  0,  t'  )  =  0.  
\eeq
The matrix $X( t )$ reads
\beq
\label{W11}
X^{A}_{\;\;B}( t )  =:  ( F,  Z^{A} ) \overleftarrow{\partial_B}( Z( t ) ).  
\eeq
The solution to the boundary problem
(\ref{W9})/(\ref{W10}) has the form
\beq
\label{W12}
\Gamma( t, t' )
=:  \frac{1}{2}  U( t )  \left[  {\rm sign}( t  -  t' )  +
\frac {  1  -  U( 1 )  }{  1  +  U( 1 )  }  \right]  U^{ -1 }( t' ),   
\eeq
where the holonomy matrix
\beq
\label{W13}
U( t )  =:  T  \exp\left\{  \int_{ 0 }^{ t }  dt'  \lambda  X( t' )  \right\}, 
\eeq
is the solution to the Cauchy problem:
\beq
\label{W14}
\partial_{ t } U( t )  =  \lambda  X( t ) U( t ),  \quad     U( t  =  0 )  =  1.  
\eeq

From  (\ref{W12})  at coincident  arguments, we have
\beq
\label{W15}
\Gamma( t, t )  =  \frac{1}{2}  U( t )  \frac{  1  -  U( 1 )  }{  1  +  U( 1
) } U^{-1}( t )
=  \frac{1}{2} 1  -  U( t ) ( 1 + U( 1 ) )^{-1}U(1)\;\! U^{-1}( t ).   
\eeq
By inserting (\ref{W15}) into (\ref{W8}), one obtains
\beq
\label{W16}
\ln J[Z]  =  -  \int_{0}^{1} d\lambda \int_{0}^{1} dt \;\! \str [  \frac {1}{2}  X( t
) -  ( 1  +  U( 1 ) )^{-1} U( 1 ) U^{-1}( t )\;\! X( t )\;\! U( t )  ].    
\eeq
On the other hand, by  differentiating (\ref{W14}) with respect to $\lambda$, we have
\beq
\label{W17}
\frac{ \partial U( 1 ) }{ \partial \lambda }  =  U( 1 )  \int_{0}^{1}  dt\;
U^{-1}( t )  X( t )  U( t ).     
\eeq
By substituting this result into the second term in the right-hand side in
(\ref{W16}), we find
\beq
\label{W18}
\ln J[Z]  =  -  \frac{1}{2}  \int _{0}^{1}  dt \;\! \str[ X( t ) ]  +  \str[  \ln ( 1
+  U( 1 ) ] | _{ \lambda  =  0 }^{  \lambda  =  1 } .  
\eeq Let us restrict the functional (\ref{W18}) on the special
trajectory (\ref{W7}). Because of (\ref{Q21}) together with
(\ref{W14})  at $\lambda  =  1$, we have \beq \label{W19}
U( 1 )|_{ \lambda  =  0 }  =  1,      
\eeq
\beq
\label{W20}
U ( 1 ) |_{ \lambda  =  1 }  =  (  Z_{R}( 1 ) \otimes \overleftarrow{\partial }
) (  Z_{R}( 0 ) \otimes \overleftarrow{\partial }  )^{-1}.    
\eeq
By using (\ref{W19}), (\ref{W20}), finally (\ref{W18}) reads
\footnote{In (\ref{W21}) and below, $\Delta$ means the standard odd Laplacian in the Darboux  co-ordinates (\ref{Q18}),
$\Delta  =  \frac{1}{2} (-1)^{ \varepsilon_{A} } \partial_{A} E^{AB} \partial_{B}.$}
\beq
\nonumber
\ln J [  Z_{R} ]   &=&    E( {\rm ad}( F ) ) (  \Delta F  )( Z_{R}( 0 ) )  -  \str \ln [ Z_{R}( 0
)\otimes\overleftarrow{\partial } ]+\\
\label{W21}
&&+   \str \ln\left[  \frac{1}{2}  (  Z_{R}( 1 )  +  Z_{R}( 0 )  ) \otimes
\overleftarrow{ \partial }  \right].    
\eeq
Due to the boundary condition (\ref{Q16}) the third term in the right-hand side in (\ref{W21})
 equals to zero.
Thus, it follows from (\ref{Q14}), (\ref{W21}) that
\beq
\nonumber
\label{W21a}
U( 1, Z,P )  &=&
\exp\left\{ -  \frac{ i }{ \hbar }  P_{A}  \left(  Z_{ R }^{A}( 1 )-Z_{ R }^{A}( 0
)  \right)\right\}\times\\
\label{W21a}
 &&\times\exp\left\{- E( {\rm ad}( F ) ) ( \Delta  F )( Z_{R}( 0 ) )  +  \str \ln[
Z_{R}( 0 ) \otimes \overleftarrow{ \partial }  ] \right\}, 
\eeq
where
\beq
\label{W21b}
Z_{R}^{A}( 0 )  =  \frac{  \exp\left\{ - \frac{1}{2}\; {\rm ad}( F ) \right\}  }
{  {\rm cosh}\left(
\frac{1}{2} \;{\rm ad}( F ) \right) } \; Z^{A}.    
\eeq
Notice also that
\beq
\nonumber
E ( {\rm ad} ( F )
) ( \Delta F )( Z_{R}( 0 ) )   &=& \frac{  {\rm sinh}\left(
\frac{1}{2} \;{\rm ad}( F ) \right) }{ \frac{1}{2}\; {\rm ad} (F)}\;
( \Delta F )\;\left({\rm cosh}^{-1}\left(  \frac{1}{2} {\rm ad}( F )  \right)  Z   \right)=\\
\label{W22}
&=&-  \frac{1}{2}  \str \ln [  Z_{ R }( 1 ) \otimes \overleftarrow{\partial }_{
Z_{ R }( 0 ) } ] ,  
\eeq where ${\rm sinh}( x ) $ is the ordinary hyperbolic sine, and
\beq \label{W23} Z_{ R }^{A}( 1 )  =  \frac{ \exp\left\{ \frac{1}{2}
\;\!{\rm ad}( F ) \right\} } { {\rm cosh}\left( \frac{1}{2}\;\! {\rm
ad}( F ) \right)} \;\! Z^{A} =  Z_{ R }^{A}( 0 )\big|_{F \rightarrow
-  F }=  \exp\{ {\rm ad}( F ) \}  Z_{ R }^{A}( 0 ).   
\eeq
Due to (\ref{W22}),  the formulae (\ref{W21}), (\ref{W21a})
become
\beq
\nonumber
\ln J[Z_{ R } ]  &=&  - \frac{1}{2}  \str \ln
[  Z_{ R }( 1 ) \otimes \overleftarrow{\partial }  ]  -  \frac{1}{2}
\str \ln [  Z_{ R }( 0 ) \otimes \overleftarrow{ \partial } ] =
\\
\nonumber
&=&  -\frac{1}{2}  \str \ln [  Z_{ R }( 0 ) \otimes \overleftarrow{
\partial } ] +  ( F \rightarrow - F )=\\
&=&  -  \frac{1}{2}  \str  \ln [  Z_{ R }( 1 )\otimes \overleftarrow{\partial }  ]
+  ( F  \rightarrow  -  F ) ,
\label{W24}
\eeq
\beq
\label{W25}
U( 1, Z, P )  =  J^{-1}[ Z_{ R } ]
\exp\left\{  -  \frac{ i }{ \hbar }  P_{A}  \left(  Z_{ R }^{A}( 1 )  -  Z_{ R }^{A}( 0
)  \right)  \right\}.  
\eeq
By inserting (\ref{Q14}) into (\ref{Q13}) one gets
\beq
\label{W26}
\Psi( 1, Z )  =
\left \langle \exp\left\{ -  \frac{ i }{ \hbar }  \int_{0}^{1}  dt  H(  Z'( t
),  P( t ) ) \right\}  \Psi_{ 0 }( Z'( 0 ) ) \right \rangle,   
\eeq
where
\beq
\label{W27}
Z'^{A}( t = 1 )  =  Z^{A}, \quad   Z'^{A}( t )  =  Z^{A}  -  \int_{t}^{1} dt'
V^{A}( t' ) .  
\eeq
Here in (\ref{W27}),  the set of integration trajectories is the same as the one
specific for the case of $ZP$-normal symbols \cite{BL},
although  the  $H( Z, P )$ in (\ref{W26}) is just the *Weyl's symbol* given by
(\ref{Q19}).  It is  also worth  to mention  that  the path
integral  (\ref{W26}) is regularized  by the condition
\beq
\label{W28}
\theta( 0 )  =  \frac{1}{2},     
\eeq appropriate to the  case of Weyl's  symbols. Thus, by comparing
(\ref{W27}) to (\ref{Q16}), one realizes that  boundary  condition
for integration trajectories  may  depend actually both on the type
of symbol  chosen,  and on the type of specific quantity which the
path-integral representation is defined for.

Due to (\ref{Q19}),  the $P$ -integration yields the delta functional (\ref{Q20}) with
$Z'_{R}$ standing for $Z_{R}$, where
\beq
\label{W29}
Z^{'A}_{ R }( t )  =  \exp\{ ( t  - 1 ) {\rm ad}( F ) \} Z^{A} ,  \quad  Z^{'A}_{ R }( 0 )
=  \exp\{ - {\rm ad}( F ) \} Z^{A}.   
\eeq The corresponding  Jacobian was calculated in \cite{BL},
formula (A.14),  up to the regularization (\ref{W28}), \beq
\label{W30} - \ln  J[ Z'_{ R } ]  =  -  \theta( 0 )  \int _{0}^{1}
dt   \str [  ( F,  Z ) \otimes \overleftarrow{\partial }  ]( Z'_{ R
}( t ) )  =
E( {\rm ad}( - F ) )\;\!\Delta F.   
\eeq
By inserting (\ref{Q20}) with $Z'_{R}$ standing for $Z_{R}$, together with (\ref{W29}),
(\ref{W30}), into (\ref{W26}), we reproduce the standard formula
\beq
\label{W31}
\Psi( 1, Z )  =  \exp\{  E( - {\rm ad}( F ) )\;\! \Delta F  \}  \Psi_{ 0 }(  \exp\{ - {\rm ad}(
F ) \}  Z   ) .   
\eeq
On the other hand, one could insert the primed trajectory (\ref{W27})/(\ref{W29}),
generated by the formula (\ref{Q13}), directly into the first line
in (4.24).  As the quantities
\beq
\label{W32}
Z^{'}_{R}(1)  =   Z,\quad    Z'_{R}( 0 )  =  \exp\{ - {\rm ad}( F ) \} Z,   
\eeq
are not related to each other with the  $F$-inversion as mapping  $F$  to  $- F$,
the second and the third equality in (\ref{W24}) does not hold
for the primed trajectory (\ref{W29}).  As to the first line in (\ref{W24}), the first
term is zero, while the second term  yields
\beq
\label{W33}
J^{-1}[ Z'_{R} ]  =  \exp\{  E( - {\rm ad}( F ) ) \Delta F( Z )  \},    
\eeq
which coincides exactly with the first exponential in (\ref{W31}).

\section{Conclusion}

In the present article we have formulated the general quantum-mechanical setting for the field-antifield  BV formalism as a hyper-gauge theory based
on the Schr\"{o}dinger equation (\ref{Q4}) with the Hamiltonian (\ref{Q5}). In terms of symmetric, Weyl's, symbols, we have resolved the equation (\ref{Q13a}) for the
symbol of the evolution operator in the form of a functional path integral (\ref{Q14}) with specific, Weyl's, boundary conditions (\ref{Q16}) for integration trajectories.
By making use of Berezin's formula (\ref{Q13}), we derive then the path-integral representation (\ref{W26}) for the wave function,  in the form of a modified path
integral with the modified boundary conditions (\ref{W27}) for primed integration trajectories.

In the simplest case of a hyper-gauge Fermion (\ref{Q19}) being a function rather than an actual operator,  we have taken the path integral (\ref{Q14}) as reduced
with the delta-functional (\ref{Q20}) concentrated on the orbit (\ref{Q21}) of an anticanonical transformation.  We have calculated explicitly the delta-functional's
Jacobian (\ref{W24}) by closed resolving  the corresponding boundary problem (\ref{W9}), (\ref{W10}) in the form (\ref{W12}). We have performed a similar reduction procedure
as applied  to the modified path integral (\ref{W26}).  In this way, we have reproduced the standard formula (\ref{W31}) as describing, together with the identification
(\ref{Q9a}),  a “canonical” part of the arbitrariness in resolving the quantum master equation (\ref{Q7})/(\ref{Q8}).  We have shown that an anticanonical transformation
encoded in (\ref{W31})  comes directly from the classic orbit equation (\ref{Q21}), while the corresponding measure, the first exponential in (\ref{W31}), comes from the
Jacobian (\ref{W24}).  In contrast to that,  in our previous article \cite{BL}, where we did use the normal $ZP$-symbols,  in the course of a similar reduction procedure,
the encoded anticanonical transformation did come together with the corresponding measure from the classical orbit equation,  while the delta-functional's  Jacobian was equal  to one.

\section*{Acknowledgments}
\noindent
I. A. Batalin would like  to thank Klaus Bering of Masaryk
University for interesting discussions.
The work of I. A. Batalin is
supported in part by the RFBR grants 14-01-00489 and 14-02-01171.
The work of P. M. Lavrov is supported by the Ministry of Education and Science of
Russian Federation, grant project 2014/387/122.

\appendix
\section*{Appendix A. Resolving equation  (3.2) for symbol of evolution\\
operator }
\setcounter{section}{1}
\renewcommand{\theequation}{\thesection.\arabic{equation}}
\setcounter{equation}{0}

The basic observation is that the equation (\ref{Q13a}) for the
symbol $U$ of the evolution operator allows for a resolution in the
variation-derivative form with respect to the external sources
$J_{A}( t ), K^{A}( t )$,
\beq
\label{A.1}
U( 1, Z, P )  =
\exp\left\{ - \frac{ i }{ \hbar } \int_{0}^{1}  dt\;\! H\left( i
\hbar \frac{ \delta }{ \delta J },  i \hbar \frac{ \delta }{ \delta
K }\right)
\right\}X( 1, Z, P ) | _{ J  =  0, K  =  0 },   
\eeq
where  $X( t, Z, P )$ resolves the following Cauchy problem
\beq
\label{A.2}
i \hbar\;\! \partial_{ t } X  =
\left[  J_{A}( t ) \left( Z^{A}  +  \frac{ i \hbar }{ 2 }
\frac{ \partial }{ \partial P_{A} }\right)
+
K^{A}( t ) \left(  P_{A}  -
\frac{ i \hbar }{ 2 }  \frac{ \partial }{ \partial Z^{A} }
(-1)^{ \varepsilon_{A} } \right)
\right]  X, \;\; X( 0 , Z, P )  =  1.  
\eeq
The latter Cauchy problem resolves explicitly
\beq
\nonumber
X( 1, Z, P )  &=&  \exp\left\{ - \frac{ i }{ \hbar } \int_{0}^{1} dt \;\! J_{A}( t )
\left[  Z^{A} +
\frac{ 1 }{2} \int_{0}^{1} dt' \;\! {\rm sign}( t - t' ) K^{A}( t' )
(-1)^{ \varepsilon_{A} }  \right]  -\right.\\
\label{A.3}
&&\qquad\left.-\frac{ i }{ \hbar } \int_{0}^{1} dt\;\! K^{A}( t ) P_{A}  \right\}.    
\eeq
By inserting the unity
\beq
\label{A.4}
1 =  {\rm const}  \int \mathcal{D} V \mathcal{D} P
\exp\left\{  \frac{ i }{ \hbar } \int_{0}^{1} dt \;\! P_{A}( t ) \;\![  V^{A}( t )  -
K^{A}( t ) (-1)^{ \varepsilon_{A} }  ]  \right\},   
\eeq
into the right-hand side in (\ref{A.1}), to the right of the first exponential, we get
\beq
\label{A.5}
U( 1, Z, P ) = \left\langle
\exp\left\{ - \frac{ i }{ \hbar } P_{A} ( Z^{A}( 1 ) - Z^{A}( 0 ) )  -
\frac{ i }{ \hbar }  \int_{0}^{1}  dt \;\! H( Z( t ), P( t ) )  \right\}
\right\rangle   ,    
\eeq
where the average and the $Z^{A}( t )$ is defined in (\ref{Q15}) and (\ref{Q17}),
respectively.

\appendix
\section*{Appendix B.  Derivation of  Berezin's formula for kernel (3.10)}
\setcounter{section}{2}
\renewcommand{\theequation}{\thesection.\arabic{equation}}
\setcounter{equation}{0}

Let  $\hat{Z}^{A}$  and  $\hat{P}_{A}$ be co-ordinate and momentum  operators
\beq
\label{B1}
{\hat Z}^{A}=:X^A,\quad
{\hat P}_{A} =:  - i  \hbar  \frac{\pa}{\pa X^A} (-1)^{\varepsilon_{A} },
\quad  [ {\hat Z}^{A},  {\hat P}_{B} ]  =  i  \hbar
\delta^{A}_{\;\;B}, \quad \varepsilon({\hat Z}^A)=\varepsilon({\hat P}_{A})
=\varepsilon_A \;\!, 
\eeq
as to apply to functions $\Psi = \Psi( X )$.  We define an arbitrary Weyl's
operator as
\beq
\label{B2}
{\hat A}=\exp\left\{{\hat Z}^{A}\frac{\pa}{\pa Z^A}+
{\hat P}_{A}\frac{\pa}{\pa P_A}\right\}A(Z,P)\Big|_{Z=0,P=0}\;\!,
\eeq
with $A( Z, P )$ being the Weyl's symbol of the operator $\hat{A}$.  By
definition,  the operator $\hat{A}$ applies to the functions $\Psi(X)$ by the rule
\beq
\label{B3}
({\hat A}\Psi)(X)=\int dY {\cal K}(X,Y) \Psi(Y)\;\!,
\eeq
in terms of the kernel  ${\cal K}( X, Y )$.  It follows from (\ref{B1}) - (\ref{B3}) that
\beq
\nonumber
{\cal K}(X,Y)&=&\exp\left\{{\hat Z}^{A}\frac{\pa}{\pa Z^A}+
{\hat P}_{A}\frac{\pa}{\pa P_A}\right\}A(Z,P)\Big|_{Z=0,P=0}\;\delta(X-Y)=\\
\label{B4}
&=&\exp\left\{
-i\hbar \frac{\pa}{\pa P_A}\frac{\pa}{\pa X^A}+
X^{A}\frac{\pa}{\pa Z^A}\right\}A(Z,P)\;\delta(X-Y)\Big|_{Z=0,P=0}\;\!.
\eeq
Due to the Baker-Campbell-Hausdorff formula, the (\ref{B4}) rewrites as
\beq
\nonumber
&&{\cal K}(X,Y)=\exp\left\{-i\hbar \frac{\pa}{\pa P_A}\frac{\pa}{\pa X^A}\right\}
\exp\left\{X^{A}\frac{\pa}{\pa Z^A}\right\}
\exp\left\{\frac{i\hbar}{2}\frac{\pa}{\pa P_A}\frac{\pa}{\pa Z^A}\right\}\times\\
\label{B5} &&\qquad\qquad\qquad\qquad \times
A(Z,P)\;\delta(X-Y)\Big|_{Z=0,P=0}\;\!.
\eeq
As the second
exponential in (\ref{B5}) applies to the right by the shift $Z
\rightarrow Z  +  X$,  the  (\ref{B5}) rewrites as
\beq \label{B6}
{\cal K}(X,Y)=\exp\left\{-i\hbar \frac{\pa}{\pa P_A}\frac{\pa}{\pa
X^A}\right\} \exp\left\{\frac{i\hbar}{2}\frac{\pa}{\pa
P_A}\frac{\pa}{\pa Z^A}\right\}
A(Z+Y,P)\;\delta(X-Y)\Big|_{Z=0,P=0}\;\!.
\eeq
with the presence of
the delta-function,  $\delta( X-Y )$, taken into account. Consider
the Fourier-integral representation for the delta-function,
\beq
\label{B7}
\delta(X-Y)=(2\pi i)^{-D}\int dK
\exp\left\{\frac{i}{\hbar}K_A(X^A-Y^A)\right\}\;\!,
\eeq
with $D$
being the number of Bosons among  $X^{A}$.  By using then the
relation
\beq
\label{B8} \!\exp\left\{-i\hbar \frac{\pa}{\pa
P_A}\frac{\pa}{\pa X^A}\right\}
\exp\left\{\frac{i}{\hbar}K_A(X^A-Y^A)\right\}\!=
\exp\left\{\frac{i}{\hbar}K_A(X^A-Y^A)\!\right\} \exp\left\{K_A
\frac{\pa}{\pa P_A}\right\}\!\!,
\eeq
and the fact that the
rightmost exponential in (\ref{B8}) applies to the right by the
shift  $P_{A} \rightarrow P_{A}  + K_{A}$, we rewrite the (\ref{B6})
in the form
\beq \nonumber
&&{\cal K}(X,Y)=(2\pi i)^{-D}\int dK \exp\left\{\frac{i}{\hbar}K_A(X^A-Y^A)\right\}\times\\
\label{B9} &&\qquad\qquad\qquad\qquad\qquad
\times\exp\left\{\frac{i\hbar}{2}\frac{\pa}{\pa P_A}\frac{\pa}{\pa
Z^A}\right\} A(Z+Y,P+K)\Big|_{Z=0,P=0}\;\!.
\eeq
Due to the
symmetric dependence of $A( Z + Y, P + K )$ on $Z, Y$  and $P, K$,
we have
\beq
\label{B10}
\exp\left\{\frac{i\hbar}{2}\frac{\pa}{\pa
P_A}\frac{\pa}{\pa Z^A}\right\} A(Z+Y,P+K)\Big|_{Z=0,P=0}=
\exp\left\{\frac{i\hbar}{2}\frac{\pa}{\pa K_A}\frac{\pa}{\pa
Y^A}\right\} A(Y,K)\;\!,
\eeq
and
\beq \label{B11}
{\cal K}(X,Y)=(2\pi
i)^{-D}\int dK \exp\left\{\frac{i}{\hbar}K_A(X^A-Y^A)\right\}
\exp\left\{\frac{i\hbar}{2}\frac{\pa}{\pa K_A}\frac{\pa}{\pa
Y^A}\right\} A(Y,K)\;\!.
\eeq
By integrating in (\ref{B11}) by
parts, we get
\beq
\label{B12}
\!\!\!\!{\cal K}(X,Y)\!=\!(2\pi
i)^{-D}\!\!\int \!dK \exp\left\{\frac{i}{\hbar}K_A(X^A-Y^A)\right\}
\exp\left\{-\frac{i\hbar}{2}\frac{\overleftarrow{\pa}}{\pa
K_A}(-1)^{\varepsilon_A} \frac{\pa}{\pa Y^A}\!\right\} A(Y,K),
\eeq
or, equivalently,
\beq
\label{B13}
\!\!\!{\cal K}(X,Y)=(2\pi i)^{-D}\int dK
\exp\left\{\frac{i}{\hbar}K_A(X^A-Y^A)\right\}
\exp\left\{\frac{1}{2}(X^A-Y^A) \frac{\pa}{\pa Y^A}\right\}A(Y,K),
\eeq
where in  (\ref{B13})  the $Y$-derivative in the second
exponential applies only to $A( Y, K )$.  Thus, we arrive at the
famous Berezin's formula \cite{Ber1}
\beq
\label{B14}
{\cal K}(X,Y)=(2\pi
i)^{-D}\int dK \exp\left\{\frac{i}{\hbar}K_A(X^A-Y^A)\right\}
A\left(\frac{1}{2}(X+Y),K\right),
\eeq
whose inverse reads
\beq
\label{B15}
A( Z, P) = \int  dX  {\cal K}\left( Z + \frac{1}{2} X,  Z - \frac{1}{2} X \right)
\exp\left\{ - \frac{ i }{ \hbar } P_A X^A \right\} .  
\eeq The star-product, $A \star B$, is defined by the formula
\beq
\label{B16} \int dY \;{\cal K}_A(X,Y)\;{\cal K}_B(Y,Z)={\cal K}_{A \star B}(X,Z)\;\!,
 \eeq
where in (\ref{B16}), we have denoted by
${\cal K}_{A}, {\cal K}_{B}, {\cal K}_{ A \star B }$ the kernel corresponding,
in the sense of (\ref{B14}), to the
symbol $A, B, A\star B$, respectively. It follows then from
(\ref{B15}), (\ref{B16}) immediately that the star-product is given
by
\beq
\nonumber
&&(A \star B)(Z,P)=\int dX\;{\cal K}_{A \star B}
\left(Z+\frac{1}{2}X,Z-\frac{1}{2}X\right)
\exp\left\{-\frac{i}{\hbar}PX\right\}=\\
\nonumber
&&\qquad\qquad\qquad
=\int dX dY\;{\cal K}_A\left(Z+\frac{1}{2}X,Y\right)\;
{\cal K}_B\left(Y,Z-\frac{1}{2}X\right)
\exp\left\{-\frac{i}{\hbar}PX\right\}=\\
\nonumber
&&\qquad\qquad\qquad\; =
(2\pi i)^{-2D}\int dX dY dQ'
dQ^{''} dP' dP^{''}
\delta\left(Q'-\frac{1}{2}\Big(Z +\frac{1}{2}X+Y\Big)\right)\times\\
\nonumber
&&\qquad\qquad\qquad\qquad\times\;\delta\left(Q^{''}-
\frac{1}{2}\Big(Y+Z-\frac{1}{2}X\Big)\right)
A(Q',P')\;B(Q^{''},P^{''})\times\\
\label{B17} &&\qquad\qquad\qquad\times
\exp\left\{\frac{i}{\hbar}\left[P'\left(Z+\frac{1}{2}X-Y\right)+
P^{''}\left(Y-Z+\frac{1}{2}X\right)-PX\right]\right\},
 \eeq
 which is
equivalent  exactly \cite{Ber1} to the standard formula (\ref{Q12})
for the $\star$. Indeed,  in (B.17),
one removes both delta-functions by taking $X$ and
$Y$-integral, and substituting
\beq
\label{B18}
X  =  2 (Q' - Q'' ), \quad  Y  =  Q'  +  Q'' -  Z,    
\eeq
so that the star-product becomes
\beq
\nonumber
&&(A \star B)(Z,P)=(2\pi i)^{-2D}\int dQ' dQ^{''} dP' dP^{''}A(Q',P')\;B(Q^{''},P^{''})\times\\
\label{B19}
&&\qquad\qquad\qquad\times
\exp\left\{\frac{2i}{\hbar}\left[(P'-P^{''})Z-P'Q^{''}+P^{''}Q'-P(Q'-Q^{''})\right]\right\},
\eeq
which is just an integral counterpart to the bi-differential operator
(\ref{Q12}).

Finally, we present here a generalization
to the Berezin's  formula (\ref{B14}) and to its inverse (\ref{B15}),
\beq
\label{B20}
\mathcal{ K }( X, Y )   =:   ( 2 \pi i )^{ - D }
\int   dK  \exp\left\{ \frac{ i }{ \hbar }  K_{A} ( X^{A}  -  Y^{A} ) \right\}
A(  \alpha  X  +  \beta  Y,  K  ),   
\eeq
\beq
\label{B21}
A( Z, P )   =   \int  dX  \mathcal{ K }(  Z  +  \beta  X,   Z  -  \alpha  X  )
\exp\left\{ - \frac{ i }{ \hbar } P_{A} X^{A} \right\},  
\eeq
where in (\ref{B20}), (\ref{B21}), parameters $\alpha, \beta$
are restricted to satisfy the condition
\beq
\label{B22}
\alpha  +  \beta  =  1.   
\eeq
The corresponding interpolating operator generalizes (\ref{B2}) in the form,
\beq
\label{B23}
\hat{ A }   =:   \exp\left\{  \hat{ Z }^{A}
\frac{ \partial }{ \partial Z^{A} }  +  \hat{ P }_{A}
\frac{ \partial }{\partial P_{A} }  +
\frac{ i  \hbar }{ 2 } ( \alpha  -  \beta ) \frac{ \partial }{ \partial P_{A} }
\frac{ \partial }{ \partial Z^{A} }  \right\}  A( Z, P ) \Big|_{ Z  =  0,  P  =  0 }. 
\eeq
For particular values of parameters we have
\begin{itemize}
\item $\alpha = 1,  \beta = 0 :  ZP$- normal form;
\item $\alpha = 0,  \beta = 1 :  PZ$- normal form;
\item $\alpha = \frac{1}{2},  \beta = \frac{1}{2}$ :  symmetric (Weyl's) form.
\end{itemize}
By making use of the same method as we did when deriving the (\ref{B19}),
one can derive from (\ref{B20}), (\ref{B21})
the corresponding star-product for symbols.

\end{document}